\documentclass[twocolumn,showpacs,superscriptaddress,prb]{revtex4-1}
\usepackage{amsfonts}
\usepackage{amsmath}
\usepackage{amssymb}
\usepackage{graphicx}

\begin{document}

\title[Coulombic effects on magnetoconductivity oscillations]{Coulombic effects on magnetoconductivity oscillations induced by
microwave excitation in multisubband two-dimensional electron systems}
\author{Yu.P. Monarkha}
\affiliation{Institute for Low Temperature Physics and Engineering, 47 Lenin Avenue, 61103
Kharkov, Ukraine}

\begin{abstract} We develop a theory of magneto-oscillations in photoconductivity of multisubband two-dimensional electron systems which takes into account strong Coulomb interaction between electrons.
In the presence of a magnetic field oriented perpendicular, internal electric fields of fluctuational origin cause fast drift velocities of electron orbit centers which affect probabilities of inter-subband scattering and the photoconductivity. For the electron system formed on the liquid helium surface, internal forces are shown to suppress the amplitude of magneto-oscillations, and change positions of magnetoconductivity minima which evolve in zero-resistance states for high radiation power.
\end{abstract}




\maketitle

\section{Introduction}

Observation of microwave-resonance-induced magnetoconductivity oscillations
in the multisubband two-dimensional (2D) electron system formed on the free
surface of liquid helium~\cite{KonKon-2009,KonKon-2010} have attracted much
interest. These oscillations appear to be very similar to $1/B$-periodic
resistivity ($R_{xx}$) oscillations observed in ultrahigh-mobility
GaAs/AlGaAs heterostructures subjected to a dc magnetic field and to strong
microwave (MW) radiation of a quite arbitrary frequency $\omega >\omega _{c}$%
~\cite{ZudSim-2001,ManSme-2002,ZudDu-2003} (here $\omega _{c}$ is the cyclotron
frequency). At low temperatures $T\simeq 0.2\,\mathrm{K}$, oscillations
reported for surface electrons (SEs) on liquid helium have a peculiar shape
which is reminiscent of a derivative of a peaky function rather than a sum
of simple maxima expected for usual scattering. With an increase in radiation power,
the minima of the magnetooscillations evolve in zero-magnetoconductivity $%
\sigma _{xx}$ states~\cite{KonKon-2010}, which are very similar to
zero-resistance states (ZRS) observed in semiconductor 2D
electron systems~\cite{ManSme-2002,ZudDu-2003}, because, under a strong magnetic field,
both $\sigma _{xx}$ and $R_{xx}$ are proportional to the effective collision
frequency $\nu (B)$. The only important difference of these two similar
phenomena is that for SEs on liquid helium magneto-oscillations are observed
only in the vicinity of a specific MW frequency $\omega =\omega _{2,1}$,
representing the resonance frequency for excitation of the second surface
subband.

In semiconductor systems, ZRS are explained~\cite{AndAle-2003} as a
consequence of the negative linear conductivity condition $\sigma _{xx}<0$
which appears for high radiation power. Due to instability of the system
under this condition, it enters a nonlinear regime and develops a steady
current state with $\sigma _{xx}(j_{0})=0$. Mechanisms of the negative
linear response conductivity of a 2D electron gas formed in semiconductor
structures are based on photon-induced impurity scattering within the ground
subband~\cite{Ryz-1969,DurSac-2003}. They were quite successful in
explaining magneto-oscillations and ZRS induced by MW radiation in
semiconductor systems. Unfortunately, they cannot be applied for explanation
of negative conductivity effects in the 2D electron system formed on the
liquid helium surface, because the MW frequency $\omega $ considered in
these theories is not restricted by the condition $\omega =\omega _{2,1}$.

The new mechanism of the negative linear response conductivity reported
recently~\cite{Mon-2011-1,Mon-2011-8} is based on nonequilibrium filling of
the second surface subband,
\begin{equation}
N_{2}>N_{1}\exp \left( -\hbar \omega _{2,1}/T_{e}\right) ,
\label{e1}
\end{equation}%
induced by MW excitation (here $N_{l}$ is the number of electrons at the
corresponding surface subband and $T_{e}$ is the electron temperature). This
condition provides a new channel for negative momentum relaxation due to
usual quasi-elastic inter-subband scattering. This possibility can be seen
already from the energy conservation for electron scattering from the
excited subband ($l=2$) to the ground subband ($l=1$).

Consider the electron energy spectrum in a magnetic filed applied
perpendicular to the surface%
\begin{equation}
\varepsilon _{l,n,X}=\Delta _{l}+\hbar \omega _{c}\left( n+1/2\right)
-eE_{\Vert }X    \label{e2}
\end{equation}%
(here $\Delta _{l}$ is the spectrum of SE states, $n=0,1,2...$ , $X$ is
the center coordinate of the cyclotron motion, and $E_{\Vert }$ is the dc
electric field directed antiparallel to the $x$-axis). Then, the energy
conservation yields
\begin{equation}
\hbar \omega _{c}\left( n-n^{\prime }\right) +\hbar \omega _{2,1}-eE_{\Vert
}\left( X-X^{\prime }\right) =0.   \label{e3}
\end{equation}%
In this equation, $\hbar \omega _{2,1}$ is not a photon quantum. It is the
energy difference for electron excitation in the $z$-direction ($\hbar
\omega _{l,l^{\prime }}=\Delta _{l}-\Delta _{l^{\prime }}$). Since $\omega
_{2,1}$ is substantially higher than $\omega _{c}$, and $eE_{\Vert }L_{B}\ll
\hbar \omega _{c}$ (here $L_{B}^2=\hbar c/eB$), scattering down the surface
levels means scattering up
the Landau levels $n^{\prime }-n\equiv m^{\ast }>0$ and we have%
\begin{equation}
eE_{\Vert }\left( X-X^{\prime }\right) =\hbar \omega _{c}\left( \frac{\omega
_{2,1}}{\omega _{c}}-m^{\ast }\right) .   \label{e4}
\end{equation}%
Thus, quasi-elastic inter-subband scattering from $l=2$ to $l=1$ will be the
scattering against the driving force ($X^{\prime }<X$), if $\omega
_{2,1}/\omega _{c}-m^{\ast }>0$, or when $B$ is a bit lower than the level
matching point $\omega _{2,1}/\omega _{c}=m^{\ast }$. At the same
conditions, electron scattering up the surface levels is obviously the
scattering along the driving force ($X^{\prime }>X$). Therefore, for
the appearance of the negative conductivity correction, the additional condition of
Eq.~(\ref{e1}) is necessary.

In semiconductor 2D electron systems~\cite{ManSme-2002}, resistance minima occur at $\omega
/\omega _{c}=m^{\ast }+1/4$. For surface electrons on
liquid helium of the areal density $n_{e}\simeq 10^{6}\,\mathrm{cm}^{-2}$, magnetoconductivity
minima reported in Ref.~\onlinecite{KonKon-2010} are also located approximately
near these "magic" numbers $m^{\ast }+1/4$. Therefore, it is interesting to
investigate if there is really such strong correlation in positions of $%
R_{xx}$ and $\sigma _{xx}$ minima for these two different phenomena induced
by MW excitation. In the single-electron theory of MW-induced
magnetooscillations~\cite{Mon-2011-8}, the conductivity minima are placed
substantially closer to the level matching numbers $m^{\ast }$, because the
Landau level broadening is very small for SEs on liquid helium. It is
expected that larger distances of $\sigma _{xx}$ minima from the level matching
points observed in the experiment will be explained by the many-electron
effect. An additional interest in studying this problem is inspired by the
observation of the resonant photovoltaic effect which emerges at the minima of
conductivity oscillations~\cite{KonCheKon-2011}.

The important conclusion of the single-electron treatment of MW-induced
magneto-oscillations is that the negative conductivity terms are large for an
electron system with extremely narrow Landau levels, when the collision
broadening of Landau levels $\Gamma _{n}$ is much smaller than temperature $%
T $. This condition is well realized for surface electrons on liquid helium.
Still, under usual experimental conditions, this electron system is in the
strong Coulomb coupling regime, which means that the average Coulomb
interaction energy of an electron $U_{C}$ is much larger than $T$ and $\hbar
\omega _{c}$. Therefore, a thorough investigation of Coulombic effects on
magnetoconductivity oscillations is required.

In this work, we report the many-electron theory of magnetoconductivity
oscillations induced by resonance MW excitation which takes into account
strong Coulomb interaction between electrons. We use the fluctuational
electric field concept~\cite{DykKha-1979} which was generalized%
~\cite{MonTesWyd-2002,MonKon-2004-book} to incorporate the collision broadening of
Landau levels. In the new theory, the Coulomb interaction between electrons
affects strongly the line-shape and broadening of magnetoconductivity
oscillations. As a result, positions of $\sigma _{xx}$ minima become
dependent strongly on electron density $n_{e}$ and on the level matching number
$m^{\ast }$. This Coulombic effect appears to be very sensitive to electron
temperature, therefore, it could be used for estimation of electron heating
in experiments on MW-induced magnetooscillations.

\section{Decay rate and occupancy of excited subbands}

At low enough temperatures, depending on electron density, the 2D electron
system formed on the free surface of liquid helium undergoes the Wigner
solid transition. This transition occurs when $U_{C}/T\simeq 137$%
~\cite{GriAda-1979}. Therefore, there is a broad range $1\ll U_{C}/T<137$, where
the electron system represents a highly correlated Coulomb liquid. These
conditions are quite usual for experiments on SEs in liquid helium. It is
well established that already at $U_{C}/T>9$ the electron velocity
autocorrelation time $\tau _{e-e}$ is very high and close to the reciprocal
of the harmonic oscillator frequency in a 2D triangular electron lattice%
~\cite{ZipBroGri-76}. This allows to simplify the transport theory, using the
assumption that SEs are at equilibrium in
the moving center-of-mass reference frame. In the semiclassical transport
theory, this simplest many-electron effect increases the effective collision
frequency of SEs by a numerical factor of the order of 2, which depends on
the perpendicular electric field $E_{\bot }$. In the presence of the
quantizing magnetic field, another kind of many-electron effect becomes
important which can substantially reduce the effective collision frequency
as compared to the result of the single-electron treatment.

The influence of Coulomb interaction on electron scattering under magnetic
field can be well understood in terms of the quasi-uniform electric field of
fluctuational origin, acting on each electron of a 2D electron liquid. It is
assumed that, under the condition $U_{C}\gg T_{e}$ , an electron is displaced
from an equilibrium position due to thermal fluctuations~\cite{DykKha-1979}
similar to an electron in a triangular Wigner lattice. When the magnetic
length $L_{B}$ is much smaller than the electron spacing $%
a $, the restoring electric field of other electrons $\mathbf{E}_{f}^{(i)}$
can be considered as a quasi-uniform field, causing a fast drift velocity of
the electron orbit center $u_{f}^{(i)}=cE_{f}^{(i)}/B$ directed
perpendicular to $\mathbf{E}_{f}^{(i)}$. The fast motion of cyclotron
obit centers affects strongly the magnetoconductivity of electrons scattered by
helium vapor atoms~\cite{LeaFozKri-1997} and capillary wave quanta (ripplons)%
~\cite{MonKon-2004-book}. Monte-Carlo simulations~\cite{FanDykLea-1997} indicate that, in
a wide range of the ratio $U_{C}/T_{e}\gg 1$, the distribution of the
fluctuational electric field is close to a Gaussian with the width parameter
given by $\sqrt{\left\langle E_{f}^{2}\right\rangle }\equiv
E_{f}^{(0)}\simeq 3\sqrt{T_{e}}n_{e}^{3/4}$.

It is remarkable that a strong quasi-uniform fluctuational field does not
smear off Landau levels, because any uniform field can be eliminated by a
proper choice of the reference frame. It is the frame where the cyclotron
orbit center is at rest, and the electron spectrum coincides with the usual
Landau spectrum. In the laboratory frame, the fluctuational field tilts
the Landau levels similar to Eq.~(\ref{e2}). We shall generalize
the many-electron theory of quantum magnetotransport in 2D Coulomb liquids developed
in Ref.~\onlinecite{MonTesWyd-2002,MonKon-2004-book} by
including in consideration inter-subband scattering. In particular,
we shall consider the decay rate of the first
excited subband, which determines subband occupancies under the MW resonance.

The ZRS of SEs are observed at low temperatures $T\leq 0.2\,\mathrm{K}$, where
electrons are predominantly scattered by capillary wave quanta. The
electron-ripplon interaction is usually described by the following
Hamiltonian~\cite{MonKon-2004-book}%
\begin{equation}
H_{\mathrm{int}}\left( \mathbf{R}_{e}\right) =\frac{1}{A}\sum_{\mathbf{q}%
}Q_{q}\left( b_{\mathbf{q}}+b_{-\mathbf{q}}^{\dag }\right) U_{q}(z_{e})e^{i%
\mathbf{q}\cdot \mathbf{r}_{e}},    \label{e5}
\end{equation}%
\begin{equation}
U_{q}\left( z\right) =\Lambda q^{2}W\left( qz\right) +eE_{\bot }-%
\frac{\partial V_{e}^{(0)}}{\partial z_{e}},    \label{e6}
\end{equation}%
\[
W\left( y\right) =\frac{1}{y^{2}}-\frac{K_{1}\left( y\right) }{y},
\]%
where $A$ is the surface area, $Q_{q}=\sqrt{\hbar q/2\rho \omega _{q}}$, $%
\omega _{q}\simeq \sqrt{\alpha /\rho }q^{3/2}$ is the ripplon spectrum, $%
\mathbf{R}_{e}=\left\{ z_{e},\mathbf{r}_{e}\right\} $, $\hbar \mathbf{q}$ is
the ripplon momentum, $b_{-\mathbf{q}}^{\dag }$ and $b_{\mathbf{q}}$ are the
creation and destruction operators, $\Lambda =e^{2}\left( \epsilon -1\right)
/4\left( \epsilon +1\right) $, $\epsilon $ is the liquid helium dielectric
constant, and $K_{1}\left( x\right) $ is the modified Bessel function of the
second kind.

In the presence of the perpendicular electric field $E_{\bot }$, the wave
functions of the first two surface subbands can be approximated by%
\begin{equation}
\varphi _{1}(z)=A_{1}ze^{-\gamma _{1}z}\mathtt{\ \ ,}\mathrm{\ }\varphi
_{2}(z)=A_{2}z\left[ 1-\frac{2\bar{\gamma}_{2,1}}{3}z\right] e^{-\gamma
_{2}z},    \label{e7}
\end{equation}%
where $A_{l}$ are normalization constants, parameters $\gamma _{l}$ are
found by the variation, and $\bar{\gamma}_{l,l^{\prime }}=\left( \gamma
_{l}+\gamma _{l^{\prime }}\right) /2$. In the magnetotransport theory, using
the dimensionless parameter $x_{q}=q^{2}L_{B}^{2}/2$, it is convenient to
represent matrix elements of Eq.~(\ref{e6}) as $\left( U_{q}\right)
_{l,l^{\prime }}=\Lambda V_{l,l^{\prime }}\left( x_{q}\right) /L_{B}^{2}$,
where
\begin{equation}
\text{ }V_{l,l^{\prime }}\left( x\right) =xw_{l,l^{\prime }}\left( \frac{x}{2%
\bar{\gamma}_{l,l^{\prime }}^{2}L_{B}^{2}}\right) +\frac{F_{l,l^{\prime
}}L_{B}^{2}}{\Lambda },    \label{e8}
\end{equation}%
\begin{equation}
F_{l,l^{\prime }}=eE_{\bot }\delta _{l,l^{\prime }}+\left( \frac{\partial
\mathrm{v}}{\partial z}\right) _{l,l}^{1/2}\left( \frac{\partial \mathrm{v}}{%
\partial z}\right) _{l^{\prime },l^{\prime }}^{1/2}-\left( \frac{\partial
\mathrm{v}}{\partial z}\right) _{l,l^{\prime }}\text{ },
\label{e9}
\end{equation}%
$\mathrm{v}(z)=-\Lambda /z+eE_{\bot }z$, and the functions $w_{l,l^{\prime
}}\left( y\right) $ represent the matrix elements of $2W\left(
qz\right) $. Useful expressions for the dimensionless functions $%
w_{l,l^{\prime }}\left( y\right) $ are given in the Appendix.

At first, consider the probability of inter-subband scattering of an
electron moving in crossed magnetic $\mathbf{B}$ and electric $\mathbf{E}%
_{\Vert }$ fields. The matrix elements of the interaction Hamiltonian yield $%
X^{\prime }-X=-q_{y}L_{B}^{2}$, therefore, the term $eE_{\Vert }\left(
X-X^{\prime }\right) $ entering the energy conservation of Eq.~(\ref{e3}) can
be rewritten as $\hbar \mathbf{q}\cdot \mathbf{u}_{d}$, where $\mathbf{u}%
_{d} $ is the drift velocity ($u_{d}=cE_{\parallel }/B$). Then, in the Born
approximation, rates of inter-subband scattering can be found as
\[
\nu _{l,n\rightarrow l^{\prime },n^{\prime }}=\frac{2\pi }{\hbar }\sum_{%
\mathbf{q}}Q_{q}^{2}\left( 2N_{q}+1\right) \left\vert \left( U_{q}\right)
_{l,l^{\prime }}\right\vert ^{2}\times
\]%
\begin{equation}
\times J_{n,n^{\prime }}^{2}\left( x_{q}\right) \delta (\varepsilon _{n%
\mathbf{^{\prime }}}-\varepsilon _{n}+\hbar \mathbf{q}\cdot \mathbf{u}%
_{d}+\Delta _{l^{\prime },l}),    \label{e10}
\end{equation}%
where $\Delta _{l,l^{\prime }}=\Delta _{l}-\Delta _{l^{\prime }}$,
\[
J_{n,n^{\prime }}^{2}(x)=\frac{[\min (n,n^{\prime })]!}{[\max (n,n^{\prime
})]!}x^{|n-n^{\prime }|}e^{-x}\left[ L_{\min (n,n^{\prime })}^{|n-n^{\prime
}|}(x)\right] ^{2},
\]%
$L_{n}^{m}(x)$ are the associated Laguerre polynomials, and $N_{q}$ is the
ripplon distribution function.

Since Eq.~(\ref{e10}) is independent of $X$, one can find the average rate $%
\bar{\nu }_{l\rightarrow l^{\prime }}$ using equilibrium electron
distribution over Landau levels: $Z_{\Vert }^{-1}e^{-\varepsilon _{n}/T_{e}}$%
, where $Z_{\Vert }=\sum_{n}e^{-\varepsilon _{n}/T_{e}}$, and $\varepsilon _{n}$ 
is the Landau spectrum. For an ensemble of
electrons, this procedure is equivalent to the assumption that electrons are
in equilibrium in the moving center-of-mass reference frame, where $\mathbf{E}%
_{\Vert }^{\prime }=0$. In order to
include the collision broadening of Landau levels, we follow Ref.%
~\onlinecite{KubMiyHas-1965} and represent the probability of scattering in
terms of density-of-state functions of the initial and final states.
Then, $\delta \left( \varepsilon -\varepsilon _{n}\right) $
of the density-of-state functions can be replaced
by $-\frac{1}{\pi \hbar }\mathrm{Im}G_{l,n}\left( \varepsilon \right) $.
Here $G_{l,n}\left( \varepsilon \right) $ is the single-electron Green's
function of the corresponding subband, whose imaginary part is broadened due
to collisions with scatterers. This procedure is equivalent to the
self-consistent Born approximation (SCBA)~\cite{AndUem-1974}.

Using the Gaussian shape of Landau level densities, inter-subband scattering
rates can be represented as
\begin{equation}
\bar{\nu}_{l\rightarrow l^{\prime }}=\frac{\hbar }{m_{e}A}\sum_{\mathbf{q}%
}\chi _{l,l^{\prime }}\left( q\right) S_{l,l^{\prime }}^{(0)}\left( q,\omega
_{l,l^{\prime }}-\mathbf{q}\cdot \mathbf{u}_{d}\right) ,\text{ \ }
\label{e11}
\end{equation}%
where $m_{e}$ is the free electron mass,
\begin{equation}
\text{\ }\chi _{l,l^{\prime }}\simeq \frac{m_{e}TL_{B}^{2}}{2\alpha \hbar
^{3}x_{q}}\left\vert \left( U_{q}\right) _{l,l^{\prime }}\right\vert ^{2},
  \label{e12}
\end{equation}%
\[
S_{l,l^{\prime }}^{(0)}\left( q,\omega \right) =\frac{2\pi ^{1/2}\hbar }{%
Z_{\parallel }}\sum_{n,n^{\prime }}\frac{J_{n,n^{\prime }}^{2}(x_{q})}{%
\Gamma _{l,n;l^{\prime },n^{\prime }}}e^{-\varepsilon _{n}/T_{e}}\times
\]%
\begin{equation}
\times \exp \left[ -\left( \frac{\hbar \omega -m\hbar \omega _{c}-\Gamma
_{l,n}^{2}/4T_{e}}{\Gamma _{l,n;l^{\prime },n^{\prime }}}\right) ^{2}+\frac{%
\Gamma _{l,n}^{2}}{8T_{e}^{2}}\right] ,    \label{e13}
\end{equation}%
$m=n^{\prime }-n$, $\Gamma _{l,n}$ is the collision broadening of Landau
levels of the corresponding subband, and $2\Gamma _{l,n;l^{\prime
},n^{\prime }}^{2}=\Gamma _{l,n}^{2}+\Gamma _{l^{\prime },n^{\prime }}^{2}$.
In Eq.~(\ref{e12}), we consider $N_{q}\simeq T/\hbar \omega _{q}\gg 1$. For
short-range scatterers, the collision broadening is independent of $n$ ($%
\Gamma =\hbar \sqrt{2\omega _{c}\nu /\pi }$, here $\nu $ is the collision
frequency in the absence of the magnetic field). In the case of
electron-ripplon interaction, the collision broadening of Landau levels
depends on $n$~\cite{MonKon-2004-book}. At $\mathbf{q}\cdot \mathbf{u}%
_{d}\rightarrow 0$, Eq.~(\ref{e11}) coincides with the decay rate found
previously in the absence of $\mathbf{E}_{\Vert }$%
~\cite{Mon-2011-8,MonSoc-2010}.

For $l=l^{\prime }$, the factor $S_{l,l^{\prime }}^{(0)}\left( q,\omega
\right) $ represents the dynamical structure factor of a nondegenerate 2D
system of noninteracting electrons. The definition of Eq.~(\ref{e13}) extends
the factor for $l\neq l^{\prime }$ to describe inter-subband scattering.
Relationship $S_{l,l^{\prime }}\left( \mathbf{q},\omega \right)
=S_{l,l^{\prime }}^{(0)}\left( q,\omega -\mathbf{q}\cdot \mathbf{u}%
_{d}\right) $, which follows from Eq.~(\ref{e11}), represents the Galilean
invariance along the interface for the generalized factor, assuming
electrons are in equilibrium in the center-of-mass reference frame, 
where $\mathbf{E}_{\Vert }^{\prime }=0$.

The Eq.~(\ref{e11}) describes how a fast drift velocity affects the
probability of inter-subband scattering. As noted above, the Coulomb liquid
can be approximately considered as an ensemble of independent electrons
whose orbit centers move fast ($\mathbf{u}_{d}\rightarrow \mathbf{u}%
_{f}^{(i)}$) due to the fluctuational electric field $\mathbf{E}_{f}^{(i)}$.
In experiments on MW-induced magneto-oscillations of surface electrons in
liquid helium, typical electron spacing $a$ varies approximately from $%
1\cdot 10^{-3}\,\mathrm{cm}$ to $3\cdot 10^{-4}\,\mathrm{cm}$, which is much
larger than the typical electron localization radius in the perpendicular
direction $\gamma _{1}^{-1}\sim 10^{-6}\,\mathrm{cm}$. Therefore, we can
assume that the fluctuational electric field acting on electrons, occupying
first few excited subbands, is the same as that acting on electrons of the
ground subband.

In the single-electron theory~\cite{Mon-2011-8}, scattering probabilities and the momentum
relaxation rate of a multisubband 2D electron system are found in terms of
the generalized dynamic factor $S_{l,l^{\prime }}^{(0)}\left( q,\omega
\right) $ using its equilibrium properties. Therefore, the
main goal of the many-electron theory is to construct the Coulomb liquid
version of this factor $S_{l,l^{\prime }}^{(\mathrm{me})}\left( q,\omega
\right) $ applicable under conditions $U_{C}/T\gg 1$. Following the ideas
developed for intra-subband scattering, we expect that the many-electron
dynamical factor of the multi-subband system of strongly interacting
electrons, can be approximated as
\begin{equation}
S_{l,l^{\prime }}^{(\mathrm{me})}\left( q,\omega \right) \simeq \left\langle
S_{l,l^{\prime }}^{(0)}\left( q,\omega -\mathbf{q}\cdot \mathbf{u}%
_{f}\right) \right\rangle _{f},   \label{e14}
\end{equation}%
where $\left\langle {...}\right\rangle _{f}$ denotes averaging over the
fluctuational electric field. For $l=l^{\prime }$, this relationship was
already proven~\cite{MonTesWyd-2002,MonKon-2004-book}. For inter-subband
scattering, the approximation of Eq.~(\ref{e14}) follows from the simple
averaging of Eq.~(\ref{e11}) over the fluctuational electric field, where $%
\mathbf{u}_{f}$ substitutes for $\mathbf{u}_{d}$. Since the maxima of $%
S_{l,l^{\prime }}^{(0)}\left( q,\omega \right) $ as a function of frequency
has a simple Gaussian shape, the model of Eq.~(\ref{e14}) provides larger
broadening of these maxima $\sqrt{\Gamma _{l,n;l^{\prime
},n^{\prime }}^{2}+x_{q}\Gamma _{C}^{2}}$ (here
$\Gamma _{C}=\sqrt{2} eE_{f}^{(0)}L_{B}$), which agrees with the broadening
of the dynamical structure factor of the 2D Wigner solid under a
strong magnetic field~\cite{Mon-2001}.

When considering inter-subband scattering, we must separate scattering down
and up the surface subbands. For electron scattering down ($l>l^{\prime }$),
at typical electron densities $n_{e}\sim 10^{6}\,\mathrm{cm}^{-2}$, the
approximation of Eq.~(\ref{e14}) is quite sufficient. For description of electron
scattering up the surface subbands, an equilibrium form of $S_{l,l^{\prime
}}^{(\mathrm{me})}\left( q,\omega \right) $ should contain also small
frequency shifts which provide the detailed balancing $\bar{\nu}_{l^{\prime
}\rightarrow l}=\bar{\nu}_{l\rightarrow l^{\prime }}\exp \left( -\hbar
\omega _{l,l^{\prime }}/T_{e}\right) $. Correct frequency shifts of $%
S_{l,l^{\prime }}^{(\mathrm{me})}\left( q,\omega \right) $ can be obtained
from the detailed balancing, or using an accurate model based on
properties of the 2D Wigner solid under a strong magnetic field%
~\cite{Mon-2001,MonKon-2004-book}. This approach yields
\[
S_{l,l^{\prime }}^{(\mathrm{me})}\left( q,\omega \right) =\frac{%
2\pi ^{1/2}\hbar }{Z_{\parallel }}\sum_{n,n^{\prime }}\frac{J_{n,n^{\prime
}}^{2}(x_{q})}{\sqrt{\Gamma _{l,n;l^{\prime },n^{\prime }}^{2}+x_{q}\Gamma
_{C}^{2}}} \times
\]
\begin{equation}
\times e^{-\varepsilon _{n}/T_{e}}I_{l,n;l^{\prime },n^{\prime }}^{(%
\mathrm{me})}\left( \omega \right) ,    \label{e15}
\end{equation}%
where
\begin{equation}
I_{l,n;l^{\prime },n^{\prime }}^{(\mathrm{me})}\left( \omega \right) =
\exp \left\{-D_{l,n;l^{\prime },n^{\prime }}(\omega )\right\},
\label{e16}
\end{equation}
\[
D_{l,n;l^{\prime },n^{\prime }}(\omega )=\frac{\left( \hbar \omega
-m\hbar \omega _{c}-\frac{ \Gamma
_{l,n}^{2}}{4T_{e}}-\frac{x_{q}\Gamma _{C}^{2}}{4T_{e}}\right) ^{2}}{\Gamma
_{l,n;l^{\prime },n^{\prime }}^{2}+x_{q}\Gamma _{C}^{2}}-\frac{\Gamma
_{l,n}^{2}}{8T_{e}^{2}},
\]
and $m=n^{\prime }-n$. Frequency shifts $\Gamma _{l,n}^{2}/4T_{e}$ and $x_{q}\Gamma
_{C}^{2} /4T_{e}$ are small as compared to $\hbar \omega _{c}$. It
should be noted that the approximation of Eq.~(\ref{e14}) gives Eqs.~(\ref%
{e15}) and (\ref{e16}) without the small frequency shift~$x_{q}\Gamma
_{C}^{2}/4T_{e}$. We restored it from the expression for the
dynamical structure factor of the Wigner solid.
This shift follows also from the detailed balancing, because
it provides the important property of the equilibrium factor%
\begin{equation}
S_{l,l^{\prime }}^{(\mathrm{me})}\left( q,-\omega \right) =e^{-\hbar \omega
/T_{e}}S_{l^{\prime },l}^{(\mathrm{me})}\left( q,\omega \right) ,
\label{e17}
\end{equation}%
which is equivalent to the\ detailed balancing.

Thus, to describe inter-subband scattering rates $\bar{\nu }_{l\rightarrow
l^{\prime }}$ in the multi-subband Coulomb liquid, one can use the general
equation similar to Eq.~(\ref{e11}), where the equilibrium many-electron
factor $S_{l,l^{\prime }}^{(\mathrm{me})}\left( q,\omega _{l,l^{\prime
}}\right) $ of Eq.~(\ref{e15}) substitutes for $S_{l,l^{\prime }}^{(0)}\left(
q,\omega _{l,l^{\prime }}-\mathbf{q}\cdot \mathbf{u}_{d}\right) $. The
Coulomb broadening of the generalized factor $\sqrt{x_{q}}\Gamma _{C}$ is
not equivalent to the collision broadening of the SCBA theory, because it
depends on the wave-vector argument of $S_{l,l^{\prime }}^{(\mathrm{me}%
)}\left( q,\omega \right) $, which complicates evaluations.

Consider the decay rate of the first excited surface subband
$\bar{\nu}_{2\rightarrow 1}$. The usage of Eqs.~(\ref{e15}) and (\ref{e16}) yields
\[
\bar{\nu}_{2\rightarrow 1}=\frac{\Lambda ^{2}T}{2\sqrt{\pi }\alpha \hbar
L_{B}^{4}Z_{\parallel }}\sum_{n,m}\frac{n!}{%
\left( n+m\right) !} e^{-\varepsilon _{n}/T_{e}}
\times
\]%
\[\times \int\limits_{0}^{\infty }V_{2,1}^{2}\left( x\right)
\frac{x^{m-1}\left[ L_{n}^{m}\left( x\right) \right] ^{2}}{\sqrt{%
\Gamma _{2,n;1,n+m}^{2}+x\Gamma _{C}^{2}}}
\times
\]
\begin{equation}
\times \exp \left[ -x-\frac{\hbar
^{2}\left( \omega _{2,1}-m\omega _{c}-x\Gamma _{C}^{2}/4\hbar T_{e}\right)
^{2}}{\Gamma _{2,n;1,n+m}^{2}+x\Gamma _{C}^{2}}\right] dx,
\label{e18}
\end{equation}%
where $m$ is a positive integer, and the frequency shift $\Gamma
_{l,n}^{2}/4T_{e}$ is disregarded as well as the small parameter
$\Gamma _{l,n}^{2}/8T_{e}^{2}$. Here we still keep the Coulomb
shift $x_{q}\Gamma _{C}^{2}/4\hbar T_{e}$ because it can be important
at high electron densities.

\begin{figure}[tbp]
\begin{center}
\includegraphics[width=10.cm]{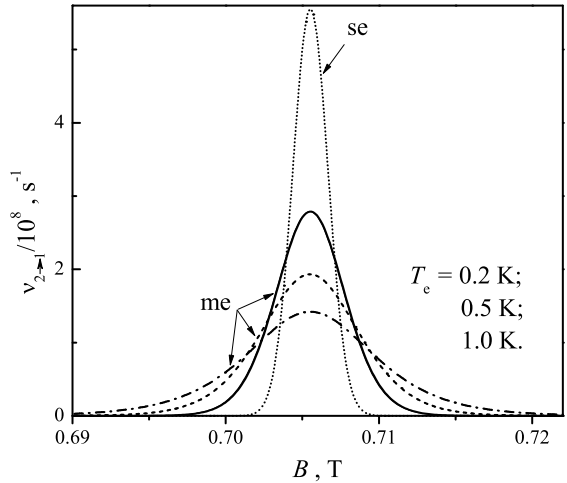}
\end{center}
\caption{Decay rate of the first excited subband $\nu _{2\rightarrow
1}$ vs the magnetic field $B$ for $T=0.2\,\mathrm{K}$ and
$n_{e}=10^{6}\,\mathrm{cm}^{-2}$: single-electron treatment (dotted);
many-electron theory [$T_{e}=T$ (solid), $T_{e}=0.5\,\mathrm{K}$ (dashed),
and $T_{e}=1\,\mathrm{K}$ (dash-dotted)].} \label{f1}
\end{figure}

For typical conditions of experiments~\cite{KonKon-2009,KonKon-2010}
performed on SEs over liquid $^{3}\mathrm{He}$, the results of numerical
evaluation of Eq.~(\ref{e18}) are shown in Fig.~\ref{f1}. The magnetic field
range is chosen to be a vicinity of the level matching number $m^{\ast }=4$.
One can see that already at a very low electron density $n_{e}=10^{6}\,\mathrm{cm%
}^{-2}$, the many-electron curve calculated for $T_{e}=T=0.2\,\mathrm{K}$
(solid line) substantially broader than the result given by the
single-electron theory (dotted line). Under these
conditions $U_{C}/T\simeq 15$. The broadening of the many-electron curve
increases strongly with $n_{e}$ and the level matching number $m^{\ast }$
due to $\Gamma _{C}\propto n_{e}^{3/4}/B^{1/2}$. Another important
conclusion, which follows from this figure, is that the many-electron curve
becomes even broader with heating of electrons because $\Gamma _{C}\propto
\sqrt{T_{e}}$. It should be noted also that the single-electron decay rate
has much weaker dependence on $T_{e}$ which is not shown in Fig.~\ref{f1}.

The above noted Coulomb broadening of the decay rate $\bar{\nu}%
_{2\rightarrow 1}$ affects strongly subband occupancies $\bar{n}_{l}=N_{l}/N$
in the presence of MW radiation. For the two-subband model, which is valid
for $T_{e}\leqslant 2K$, we have
\begin{equation}
\frac{\bar{n}_{2}}{\bar{n}_{1}}=\frac{1+\exp \left( -\hbar \omega
_{2,1}/T_{e}\right) \bar{\nu}_{2\rightarrow 1}/r_{\mathrm{mw}}}{1+\bar{\nu}%
_{2\rightarrow 1}/r_{\mathrm{mw}}},    \label{e19}
\end{equation}%
where $r_{\mathrm{mw}}$ is the MW excitation rate. Under the resonance
condition, $r_{\mathrm{mw}}=\Omega _{R}^{2}/2\gamma _{\mathrm{mw}}$, where $%
\gamma _{\mathrm{mw}}$ is the half-width of the resonance, and $\Omega _{R}$
is the Rabi frequency proportional to the amplitude of the MW field. Thus,
the fluctuational electric field leads to much broader variations of $\bar{n}%
_{2}(B)$ near the level matching condition $\omega _{2,1}/\omega
_{c}=m^{\ast }$ than those given by the single-electron theory.

\begin{figure}[tbp]
\begin{center}
\includegraphics[width=10.cm]{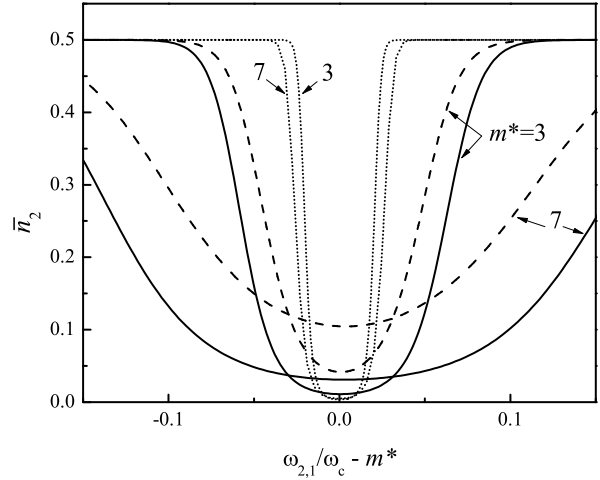}
\end{center}
\caption{The occupancy of the first excited subband vs the parameter $\omega _{2,1}/\omega
_{c}-m^{\ast }$ for $T=0.2\,\mathrm{K}$,
$n_{e}=2\cdot 10^{6}\,\mathrm{cm}^{-2}$: single-electron
treatment (dotted), and many-electron theory [$\Omega _{R}=2.5\cdot 10^{7}\,\mathrm{s}^{-1}$
(solid), and $\Omega _{R}=5\cdot 10^{7}\,\mathrm{s}^{-1}$ (dashed)] .} \label{f2}
\end{figure}

For electron density $n_{e}=2\cdot 10^{6}\,\mathrm{cm}^{-2}$ and $T_{e}=T=0.2%
\,\mathrm{K}$, variations of the fractional occupancy $\bar{n}_{2}$ are shown
in Fig.~\ref{f2}. When conducting numerical evaluations, we assume that the
width of the MW resonance is due to inhomogeneous broadening $2\gamma _{%
\mathrm{mw}}=0.3\,\mathrm{GHz}$~\cite{KonKon-2010}, and the MW excitation rate $%
r_{\mathrm{mw}}$ is independent of $B$. The many-electron lines calculated
for $m^{\ast }=3$ and $7$ at $\Omega _{R}=2.5\cdot 10^{7}\,\mathrm{s}^{-1}$
(solid) are substantially broader than the corresponding single-electron
lines (dotted). The increase of $\bar{n}_{2}$ with MW power is shown by
many-electron lines (dashed) calculated at $\Omega _{R}=5\cdot 10^{7}\,\mathrm{%
s}^{-1}$. Since the negative conductivity correction is proportional to $%
\bar{n}_{2}-\bar{n}_{1}\exp \left( -\hbar \omega _{2,1}/T_{e}\right) $, it
is clear that the many-electron effect should displace strongly
conductivity minima.

\section{Linear magnetoconductivity}

According to the analysis given in the previous Section, probabilities of
electron scattering in a multi-subband Coulomb liquid ($U_{C}/T>10$) can be
described by the many-electron dynamic factor $S_{l,l^{\prime }}^{(\mathrm{me%
})}\left( q,\omega \right) $ of Eq.~(\ref{e15}) instead of $S_{l,l^{\prime
}}^{(0)}\left( q,\omega \right) $ of the single-electron theory. Therefore,
using equilibrium properties of $S_{l,l^{\prime }}^{(\mathrm{me})}\left(
q,\omega \right) $, like that given in Eq.~(\ref{e17}), and following the
formal procedure described in details in Ref.~\onlinecite{Mon-2011-8}, one can
obtain an equation for the momentum relaxation rate similar to that found in
the single-electron treatment. The effective collision frequency $\nu $\ is
a sum of contributions from intra-subband ($\nu _{\mathrm{intra}}$) and
inter-subband ($\nu _{\mathrm{inter}}$) scattering. The $\nu _{\mathrm{inter}%
}$ consists of normal terms ($\nu _{\mathrm{N}}$), existing even at $\bar{n}%
_{l}=\bar{n}_{l^{\prime }}\exp \left( -\hbar \omega _{l,l^{\prime
}}/T_{e}\right) $, and anomalous terms $\nu _{\mathrm{A}}$ proportional to $%
\bar{n}_{l}-\bar{n}_{l^{\prime }}\exp \left( -\hbar \omega _{l,l^{\prime
}}/T_{e}\right) $. Simple replacement $S_{l,l^{\prime }}^{(0)}\left(
q,\omega \right) \rightarrow S_{l,l^{\prime }}^{(\mathrm{me})}\left(
q,\omega \right) $ yields
\begin{equation}
\nu _{\mathrm{intra}}=\frac{\hbar \omega _{c}^{2}}{4\pi T_{e}}\sum_{l}\bar{n}%
_{l}\int\limits_{0}^{\infty }x_{q}\chi _{l,l}\left( q\right) S_{l,l}^{(%
\mathrm{me})}\left( q,0\right) dx_{q},    \label{e20}
\end{equation}%
\[
\text{\ }\nu _{\mathrm{N}}=\frac{\hbar \omega _{c}^{2}}{4\pi T_{e}}%
\sum_{l>l^{\prime }}\left( \bar{n}_{l}+\bar{n}_{l^{\prime }}e^{-\Delta _{l,l^{\prime
}}/T_{e}}\right) \times
\]%
\begin{equation}
\times \int_{0}^{\infty }x_{q}\chi _{l,l^{\prime }}\left( q\right)
S_{l,l^{\prime }}^{(\mathrm{me})}\left( q,\omega _{l,l^{\prime }}\right)
dx_{q},    \label{e21}
\end{equation}%
\[
\nu _{\mathrm{A}}=-\frac{\omega _{c}^{2}}{4\pi }\sum_{l>l^{\prime }}\left[
\bar{n}_{l}-\bar{n}_{l^{\prime }}e^{-\Delta _{l,l^{\prime }}/T_{e}}\right] \times
\]
\begin{equation}
\times \int_{0}^{\infty }x_{q}\chi _{l,l^{\prime }}\left( q\right) \Psi
_{l,l^{\prime }}\left( x_{q}\right) dx_{q},    \label{e22}
\end{equation}%
were
\begin{equation}
\Psi _{l,l^{\prime }}\left( x_{q}\right) =-S_{l,l^{\prime }}^{\prime (%
\mathrm{me})}\left( q,\omega _{l,l^{\prime }}\right) +e^{\Delta
_{l,l^{\prime }}/T_{e}}S_{l^{\prime },l}^{\prime (\mathrm{me})}\left(
q,-\omega _{l,l^{\prime }}\right) ,    \label{e23}
\end{equation}%
and $S_{l,l^{\prime }}^{\prime (\mathrm{me})}\left( q,\omega \right) \equiv
\partial S_{l,l^{\prime }}^{(\mathrm{me})}\left( q,\omega \right) /\partial
\omega $. The expression for $\nu _{\mathrm{A}}$ was represented in the form
of Eq.~(\ref{e22}), containing terms with $l>l^{\prime }$ only, by
interchanging the running indices. The similar procedure resulted in Eq.~(\ref%
{e21}) employs the property of Eq.~(\ref{e17}).

Using the relationship
\begin{equation}
S_{l^{\prime },l}^{\prime (\mathrm{me})}\left( q,-\omega \right) =-e^{-\frac{%
\hbar \omega }{T_{e}}}S_{l,l^{\prime }}^{\prime (\mathrm{me})}\left(
q,\omega \right) +\frac{\hbar }{T_{e}}e^{-\frac{\hbar \omega }{T_{e}}%
}S_{l,l^{\prime }}^{(\mathrm{me})}\left( q,\omega \right) ,
\label{e24}
\end{equation}%
which follows from Eqs.~(\ref{e15})-(\ref{e17}), the expression for $\Psi
_{l,l^{\prime }}\left( x_{q}\right) $ can be rearranged to a more convenient form:
\[
\Psi _{l,l^{\prime }}\left( x_{q}\right) =\frac{8\pi ^{1/2}\hbar ^{2}}{%
Z_{\parallel }}\sum_{n,n^{\prime }}\frac{\left\vert J_{n,n^{\prime
}}(x_{q})\right\vert ^{2}}{\sqrt{\Gamma _{l,n;l^{\prime },n^{\prime
}}^{2}+x_{q}\Gamma _{C}^{2}}}e^{-\varepsilon _{n}/T_{e}}\times
\]%
\begin{equation}
\times I_{l,n,l^{\prime };n^{\prime }}^{(\mathrm{me})}\left( \omega
_{l,l^{\prime }}\right) \frac{\hbar \left( \omega _{l,l^{\prime }}-
m \omega _{c}\right) + \left( \Gamma _{l^{\prime
},n^{\prime }}^{2}-\Gamma _{l,n}^{2}\right) /8T_{e}}{\Gamma _{l,n;l^{\prime
},n^{\prime }}^{2}+x_{q}\Gamma _{C}^{2}},   \label{e25}
\end{equation}%
where $m=n^{\prime }-n$. It is important that the Coulomb shift
$x_{q}\Gamma _{C}^{2}/T_{e}$ does not
enter the numerator of Eq.~(\ref{e25}), where $\left( \Gamma _{l^{\prime
},n^{\prime }}^{2}-\Gamma _{l,n}^{2}\right) /8T_{e}$ can be disregarded for
extremely narrow Landau levels. Therefore, the vanishing point of each term
entering $\nu _{\mathrm{A}}(B)$ is unaffected by the many-electron effect
and practically coincides with the level matching condition $\omega
_{2,1}=\left( n^{\prime }-n\right) \omega _{c}$. At the same time, positions
of minima and maxima of $\nu _{\mathrm{A}}\left( B\right) $ are very
sensitive to electron density due to the broadening of the generalized
dynamic factor $\sqrt{\Gamma _{l,n;l^{\prime },n^{\prime }}^{2}+x_{q}\Gamma
_{C}^{2}}$. Additionally, at high $n_{e}$, they are slightly affected by the
frequency shift $x_{q}\Gamma _{C}^{2}/4T_{e}$ entering $I_{l,n,l^{\prime
};n^{\prime }}^{(\mathrm{me})}\left( \omega _{l,l^{\prime }}\right) $.

\begin{figure}[tbp]
\begin{center}
\includegraphics[width=10.cm]{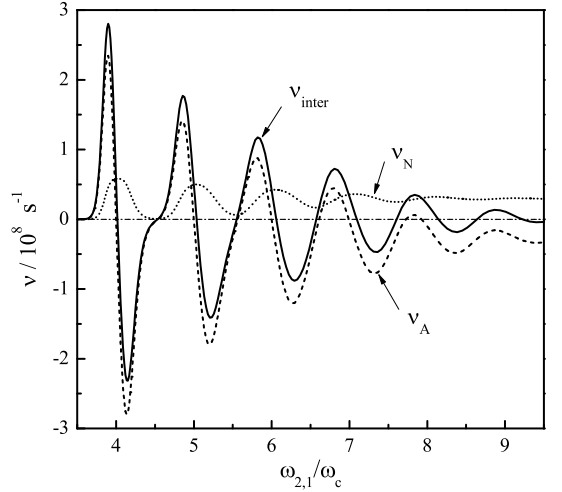}
\end{center}
\caption{The effective collision frequency due to inter-subband scattering
$\nu _{\mathrm{inter}}$ vs the
parameter $\omega _{2,1}/\omega _{c}$ (solid). Two contributions
$\nu _{\mathrm{N}}$ and $\nu _{\mathrm{A}}$ are shows by dotted and
dashed lines respectively.  } \label{f3}
\end{figure}

Typical behavior of $\nu _{\mathrm{inter}}$ and its components ($\nu _{%
\mathrm{A}}$ and $\nu _{\mathrm{N}}$) as functions of the parameter
$\omega _{2,1}/\omega _{c}$ is shown in Fig.~\ref{f3} for
sufficiently high electron density ($n_{e}=1.5\cdot 10^{7}\,\mathrm{cm}^{-2}$,
$\Omega _{R}=5\cdot 10^{7}\,\mathrm{s}^{-1}$), when maxima of $S_{l,l^{\prime
}}^{(\mathrm{me})}\left( q,\omega \right) $ begin to overlap due to the
Coulomb broadening $\sqrt{x_{q}}\Gamma _{C}$. Oscillations of $\nu _{\mathrm{%
A}}(B)$ and $\nu _{\mathrm{N}}(B)$ are quite different. The normal
contribution $\nu _{\mathrm{N}}$ is just a sum of simple peaks centered at
the level matching conditions $\omega _{2,1}=m^{\ast }\omega _{c}$. With an
increase of the parameter $\omega _{2,1}/\omega _{c}$, maxima overlap strongly, and
eventually the dependence $\nu _{\mathrm{N}}(B)$ acquires a positive
background.

The sign-changing correction $\nu _{\mathrm{A}}$ originates from
the derivative $S_{l,l^{\prime }}^{\prime (\mathrm{me})}\left( q,\omega
\right) $, and, in the vicinity of the level matching conditions, it is
close to zero, if overlapping is small. At larger $m^{\ast }$, the
overlapping of maxima of $I_{l,n,l^{\prime };n^{\prime }}^{(\mathrm{me}%
)}\left( \omega _{l,l^{\prime }}\right) $ results in a negative background
of $\nu _{\mathrm{A}}(B)$, because the amplitude of $\nu _{\mathrm{A}}$
oscillations decreases with $m^{\ast }$, and, at a fixed $B$, the negative
contribution of smaller $m^{\ast }$ dominates. It is remarkable that the
negative background of $\nu _{\mathrm{A}}(B)$ is compensated by the positive
background of $\nu _{\mathrm{N}}(B)$ and the entire contribution from
inter-subband scattering $\nu _{\mathrm{inter}}=$ $\nu _{\mathrm{N}}+\nu _{%
\mathrm{A}}$ (solid line) oscillates nearly with the zero background.

\begin{figure}[tbp]
\begin{center}
\includegraphics[width=10.cm]{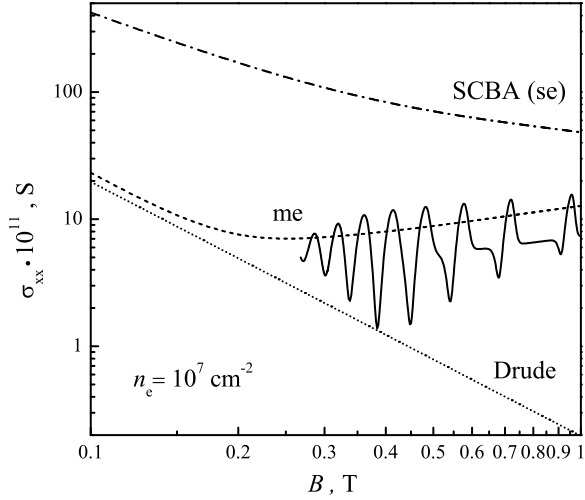}
\end{center}
\caption{Magnetoconductivity vs $B$ for $T=0.2\,\mathrm{K}$, and
$n_{e}= 10^{7}\,\mathrm{cm}^{-2}$: the single-electron treatment based on the SCBA
(dash-dotted), the Drude approximation (dotted), and the many-electron theory
(dashed). The result obtained for resonant MW excitation with $\Omega _{R}=3.5\cdot
10^{7}\,\mathrm{s}^{-1}$ is shown by the solid line.} \label{f4}
\end{figure}

The conductivity equation, which follows from the momentum-balance equation method%
~\cite{MonKon-2004-book}, has the usual Drude form, where, instead of the
quasiclassical collision frequency $\nu _{0}$ independent of $B$, it is
necessary to use the effective collision frequency $\nu =\nu _{\mathrm{intra}%
}+\nu _{\mathrm{N}}+\nu _{\mathrm{A}}$. Therefore, under strong magnetic
fields, $\sigma _{xx}$ is
proportional to $\nu _{\mathrm{intra}}+\nu _{\mathrm{N}}+\nu _{\mathrm{A}}$.
Without MW radiation, the influence of the strong Coulomb interaction on $%
\sigma _{xx}$ is illustrated in Fig.~\ref{f4} for $n_{e}=10^{7}\,\mathrm{cm}%
^{-2}$ and $T=0.2\,\mathrm{K}$ ($U_{C}/T\simeq 47$; liquid $^{3}\mathrm{He}$).
The result of the single-electron treatment based on the SCBA theory
(dash-dotted line) was obtained from Eq.~(\ref{e20}) by setting $\bar{n}%
_{1}\rightarrow 1$ and $\Gamma _{C}\rightarrow 0$. When lowering the
magnetic field, the many-electron line (dashed) moves from the SCBA line to
the Drude line ($\nu \rightarrow \nu _{0}$, dotted line). Transition to the
Drude line occurs due to the Coulomb broadening of $I_{1,n;1,n^{\prime }}^{(%
\mathrm{me})}\left( 0\right) $ which triggers off scattering processes with $%
\left\vert n^{\prime }-n\right\vert >0$. The many-electron line of Fig.%
~\ref{f4} takes into account terms with $\left\vert n^{\prime }-n\right\vert \leq
9$.

In the presence of MW radiation, typical behavior of $\sigma _{xx}(B)$ is
shown in Fig.~\ref{f4} by the solid line calculated for $\Omega _{R}=3.5\cdot
10^{7}\,\mathrm{s}^{-1}$, $n_{e}=10^{7}\,\mathrm{cm}^{-2}$, and for a restricted
field range covering the level-matching numbers $3\leq m^{\ast }\leq 9$.
This line illustrates evolution of oscillations and conductivity minima with
lowering $B$. In contrast to the single-electron theory, where the amplitude
of conductivity oscillations increases with $m^{\ast }$ (lowering $B$)%
~\cite{Mon-2011-1,Mon-2011-8}, here the amplitude of oscillations decreases with $%
m^{\ast }$ for sufficiently low magnetic fields, due to the Coulomb broadening
of $I_{2,n,1;n+m^{\ast }}^{(\mathrm{me})}\left( \omega _{2,1}\right) $. In
accordance with Fig.~\ref{f3}, conductivity minima of Fig.~\ref{f4} are
quite distant from the level matching points.

\begin{figure}[tbp]
\begin{center}
\includegraphics[width=10.cm]{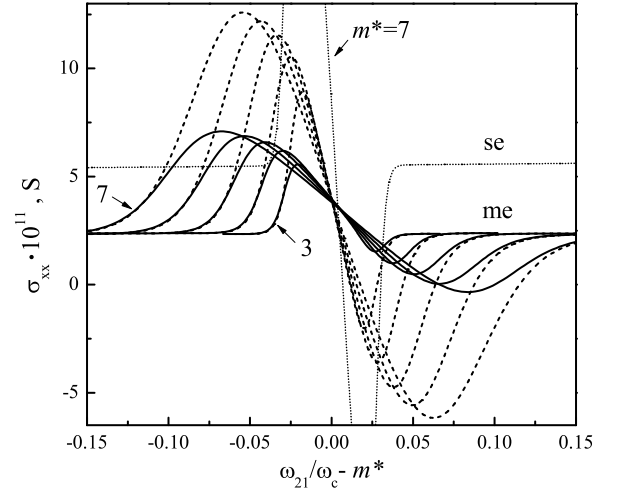}
\end{center}
\caption{Magnetoconductivity vs the parameter $\omega _{2,1}/\omega
_{c}-m^{\ast }$ for $T=0.2\,\mathrm{K}$, and
$n_{e}= 10^{6}\,\mathrm{cm}^{-2}$: the single-electron treatment (dotted),
the many-electron theory [$\Omega _{R}=2.5\cdot 10^{7}\,\mathrm{s}^{-1}$
(solid), and $\Omega _{R}=5\cdot 10^{7}\,\mathrm{s}^{-1}$ (dashed)].} \label{f5}
\end{figure}

The dependence of positions of conductivity minima on the level matching
number~$m^{\ast }$ is illustrated in Fig.~\ref{f5} for lower density $%
n_{e}=10^{6}\,\mathrm{cm}^{-2}$ and two MW powers. This figure indicates that
the distance of conductivity extremes of the many-electron lines
(solid) from the point $\omega _{2,1}/\omega _{c}=m^{\ast }$ increases fast
with $m^{\ast }$. For the comparison reason, the corresponding
single-electron line (dotted) of $m^{\ast }=7$ is also plotted. It is
interesting to note that under the assumption $T_{e}=T$ , an increase in
power (see dashed lines) reduces a little bit the distance of $\sigma _{xx}$
extremes from the level matching points. This behavior correlates with the
behavior of corresponding $\bar{n}_{2}$ lines in Fig.~\ref{f2}. The electron
heating induced by MW radiation acts in the opposite way, increasing the
Coulomb broadening of $I_{l,n,l^{\prime };n^{\prime }}^{(\mathrm{me})}\left(
\omega _{l,l^{\prime }}\right) $ (the frequency shift $x_{q}\Gamma
_{C}^{2}/4T_{e}$ is independent of $T_{e}$) due to $\Gamma _{C}\propto \sqrt{%
T_{e}}$. Therefore, positions and broadening of the conductivity minima in
an experiment can be used for estimation of electron temperature realized in
this system in the presence of resonant radiation.

\begin{figure}[tbp]
\begin{center}
\includegraphics[width=10.cm]{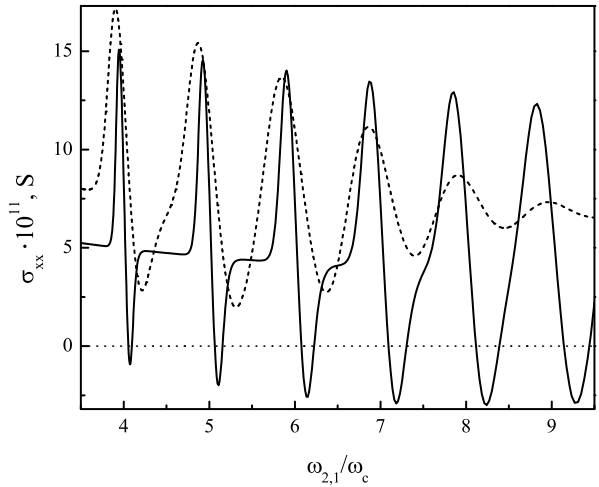}
\end{center}
\caption{Magnetoconductivity vs $\omega _{2,1}/\omega
_{c}$ for $T=0.2\,\mathrm{K}$ and two electron densities: $n_{e}= 5\cdot 10^{6}\,\mathrm{cm}^{-2}$
(solid) and $n_{e}= 2\cdot 10^{7}\,\mathrm{cm}^{-2} $ (dashed).} \label{f6}
\end{figure}

Positions of $\sigma _{xx}$ extremes depend also on $n_{e}$, as
illustrated in Fig.~\ref{f6} representing results obtained for two electron densities: $5\cdot 10^{6}\,\mathrm{cm}^{-2}$,
and $2\cdot $ $10^{7}\,\mathrm{cm}^{-2}$. One can see differences in
variations of conductivity minima and maxima. Conductivity minima definitely become
more distant from the level matching points with the increase in
$n_{e}$ for all $m^{\ast }$ presented in the figure. Regarding conductivity
maxima, with the increase in $n_{e}$, they become more distant for three
smallest level matching numbers $m^{\ast }=4,5$ and $6$, while for larger
numbers ($m^{\ast }=8$ and $9$) they become even closer to the level matching points.
For the transition number $m_{c}^{\ast }=7$, the position of the corresponding maximum is
practically unchanged. The $m_{c}^{\ast }$ becomes larger, if lower electron densities are
compared.

\section{Discussion and conclusions}

The expression for the generalized dynamic factor given in Eqs.~(\ref{e15}) and
(\ref{e16}) together with equations for the effective collision frequency
[Eqs.~(\ref{e20})-(\ref{e22})] allows to describe photoconductivity of SEs on
liquid helium in a wide range of the parameter $U_{C}/T$ up to the Wigner solid
transition. The main conclusion of the many-electron theory reported in this
work is that the unusual MW-induced magneto-oscillations of $\sigma _{xx}$
and ZRS can exist in multisubband 2D electron systems even under strong
Coulomb coupling conditions ($U_{C}\gg T$) relevant to experiments%
~\cite{KonKon-2010}.

Coulomb forces acting between electrons affect strongly the broadening of
magneto-oscillations and positions of conductivity extremes, which become
dependent on electron density, the level matching number $m^{\ast }$ and the
electron temperature. Therefore, for SEs on liquid helium, in the
many-electron theory, there are no fixed "magic" numbers describing
positions of conductivity minima on the $\omega _{2,1}/\omega _{c}$-axis.
The proximity of experimentally observed minima to $\omega /\omega _{c}\simeq
m^{\ast }+1/4$, most likely, is accidental (valid only for a specific
electron density). A substantial dependence of the distance of $\sigma _{xx}$
minima from the point $\omega /\omega _{c}=m^{\ast }$ on the level matching
number $m^{\ast }$, which can be seen in experimental data of Ref.%
~\onlinecite{KonKon-2010}, agrees qualitatively with this conclusion.

Numerical calculations of SE magnetoconductivity shown in the previous
Section were performed under the assumption that electrons are not
substantially heated by the MW ($T_{e}\simeq T$). This assumption allows the
many-electron theory to reproduce remarkable shapes of $\sigma _{xx}$
oscillations observed for SEs on liquid helium. The inclusion of Coulomb
interaction in the theory affects crucially the dependence of the amplitude
of magnetooscillations on the magnetic field strength. In the many-electron
theory, the amplitude of oscillations eventually begins decreasing with
lowering $B$, which agrees with experimental observations. Even though the
Coulombic effect suppresses amplitudes of magneto-oscillations, under
experimental conditions of Ref.~\onlinecite{KonKon-2010}, the linear theory results
in $\sigma _{xx}<0$ at certain ranges of magnetic and MW fields.

The comparison of theoretical and experimental $\sigma _{xx}(B)$ curves
indicates that typically the real heating of SEs is not strong and most likely $%
T_{e}\sim T$. Since $\nu _{\mathrm{intra}}\sim 1/T_{e}(\Gamma
_{l,n;l^{\prime },n^{\prime }}^{2}+x_{q}\Gamma _{C}^{2})^{1/2}$, a strong
increase in electron temperature ($T_{e}\gg T$) would lead to deep drops of $%
\sigma _{xx}$ at the both sides of the level matching conditions, which
contradicts to observations. In the single-electron theory, with an increase
in electron temperature the sign-changing correction $\nu _{\mathrm{A}}$
becomes more efficient than $\nu _{\mathrm{intra}}$ and $\nu _{\mathrm{N}}$,
because $\Gamma _{l,n}$ is independent of $T_{e}$. In the many-electron
theory, the Coulomb correction to the broadening of $S_{l,l^{\prime }}^{(%
\mathrm{me})}\left( q,\omega _{l,l^{\prime }}\right) $ increases with
electron temperature, $\sqrt{x_{q}}\Gamma _{C}\propto T_{e}^{1/2}$, and the\
relative increase in efficiency of $\nu _{\mathrm{A}}$ as compared to $\nu _{%
\mathrm{intra}}$ and $\nu _{\mathrm{N}}$ approaches a saturation, when $%
\sqrt{x_{q}}\Gamma _{C}\gg \Gamma _{l,n;l^{\prime },n^{\prime }}$. The
electron heating affects positions of $\sigma _{xx}$ extremes, which also
can be used for estimation of $T_{e}$. The many-electron theory of cold SEs
results in minima distances from the level matching points which
are of the same order, as those observed experimentally. This also
agrees with the estimation $T_{e}\sim T$ given above.

Regarding negative conductivity values obtained in the linear theory,
regions with $\sigma _{xx}<0$ are surely unstable, because any electron
density fluctuation produces an electric field which increases the fluctuation.
For the Coulomb liquid formed on the liquid helium surface, even the
fluctuational field $\mathbf{E}_{f}^{(i)}$ can be a cause for instability,
because excited SEs will be scattered against the restoring force $-e\mathbf{%
E}_{f}^{(i)}$. The edge of the 2D electron system on liquid helium is
usually fixed by application of a dc confining field of Corbino and guard electrodes
which also can be a cause for instability of the system at $\sigma _{xx}<0$.

A stable state can be achieved under the nonlinear transport regime by forming
a strong steady current $j_{0}$, of which $\sigma _{xx}(j_{0})=0$. In the
2D Coulomb liquid on the surface of liquid helium, formation of current
domains is unlikely because of strong electron correlations.
The SEs usually have no source and drain electrodes.
Therefore, a steady current can be formed only
by electrons circling the center of the electron pool. Since it is impossible to
create a strong current density in the center, electrons will move to the edges of the
electron liquid depleting the center. As a result, a nonuniform electron
density distribution along the surface will be formed to provide radial
electric field and a circling current, strong enough to make
$\sigma _{xx}=0$. At a fixed magnetic field, a change in electron density additionally
helps the system to leave the unstable regime, due to the many-electron
effect described here.

Thus, under the condition $\sigma _{xx}<0$ of the linear transport regime,
electron moving against the confining force of Corbino and guard electrodes
is the most likely way to reach a stable state.
This scenario agrees with recent observations of the
resonant photovoltaic effect for SEs on liquid helium~\cite{KonCheKon-2011},
where MW excitation causes a strong displacement of surface electrons
towards the edges (against the confining force), if the magnetic field is
close to the minima of the conductivity oscillations. Therefore,
experimental observation of the strong density dependence of positions of
conductivity minima consistent with the results of the many-electron theory
would be an additional evidence for the negative conductivity as the origin
of the resonant photovoltaic effect.


\section{Appendix}

According to definitions given in Eqs.~(\ref{e6})-(\ref{e8}), for the
two-subband model ($l=1,2$), the functions $w_{l,l^{\prime }}\left( y\right)
$, describing electron-ripplon coupling, can be written as%
\begin{equation}
w_{l,l^{\prime }}\left( y\right) =\frac{A_{l}A_{l^{\prime }}}{4\bar{\gamma}%
_{l,l^{\prime }}^{3}}\int_{0}^{\infty }e^{-x}\left[ \frac{1}{y}-\frac{x}{%
\sqrt{y}}K_{1}(x\sqrt{y})\right] \lambda _{l,l^{\prime }}\left( x\right) dx,
  \label{e26}
\end{equation}%
where $\lambda _{1,1}=1$, and%
\[
\lambda _{2,1}\left( x\right) =\left( 1-\frac{1}{3}x\right) ,\text{ \ }%
\lambda _{2,2}\left( x\right) =\left[ 1-\frac{\bar{\gamma}_{2,1}}{3\gamma
_{2}}x\right] ^{2}.
\]%
The integrals of the right part of Eq.~(\ref{e26}) can be found analytically,
and we have $w_{1,1}\left( y\right) =w_{1}(y)$,
\[
w_{2,1}\left( y\right) =\frac{A_{2}A_{1}}{4\bar{\gamma}_{2,1}^{3}}\left[
w_{1}(y)-\frac{1}{3}w_{2}(y)\right] ,
\]%
\[
w_{2,2}\left( y\right) =\frac{A_{2}^{2}}{4\gamma _{2}^{3}}\left[ w_{1}(y)-%
\frac{2\bar{\gamma}_{2,1}}{3\gamma _{2}}w_{2}(y)+\left( \frac{\bar{\gamma}%
_{2,1}}{3\gamma _{2}}\right) ^{2}w_{3}(y)\right] .
\]%
Here we use the following notations%
\[
w_{1}(y)=-\frac{1}{1-y}+\frac{1}{\left( 1-y\right) ^{3/2}}\ln \left[ \frac{1+%
\sqrt{1-y}}{\sqrt{y}}\right] ,
\]%
\[
w_{2}(y)=-\frac{4-y}{(1-y)^{2}}+\frac{3}{\left( 1-y\right) ^{5/2}}\ln \left[
\frac{1+\sqrt{1-y}}{\sqrt{y}}\right] ,
\]%
\[
w_{3}(y)=-\frac{19-6y+2y^{2}}{(1-y)^{3}}+\frac{(12+3y)}{\left( 1-y\right)
^{7/2}}\ln \left[ \frac{1+\sqrt{1-y}}{\sqrt{y}}\right] .
\]%
The analytical presentation of $w_{l,l^{\prime }}\left( y\right) $ given above
is very useful for numerical evaluations of $\sigma _{xx}$.

\end{document}